\documentclass[preprint]{elsarticle}
\usepackage[a4paper, total={7in, 8in}]{geometry}
\usepackage[utf8]{inputenc}
\usepackage{floatpag}
\usepackage{amsthm}
\usepackage{enumerate}
\usepackage{amssymb}
\usepackage{amsmath}
\usepackage{braket}
\usepackage{hyperref}
\hypersetup{colorlinks = true, linkcolor=blue }

\usepackage{graphicx}
\usepackage{epstopdf}
\usepackage{color} 
\graphicspath{{./Figures/}}
\usepackage{caption}
\usepackage{subcaption}
\usepackage{cleveref}

\journal{Physics Letters B}

\begin{document}

\begin{frontmatter}

\title{{\bf Finite width effects on the azimuthal asymmetry in proton-nucleus collisions in the Color Glass Condensate}}

\author[a,b]{Pedro Agostini}
\ead{pedro.agostini@usc.es}
\author[a]{Tolga Altinoluk}
\ead{tolga.altinoluk@ncbj.gov.pl}
\author[b]{N\'estor Armesto}
\ead{nestor.armesto@usc.es}

\address[a]{Theoretical Physics Division, National Centre for Nuclear Research, Pasteura 7, Warsaw 02-093,
Poland}
\address[b]{Instituto Galego de F\'{\i}sica de Altas Enerx\'{\i}as IGFAE, Universidade de Santiago de Compostela, 15782 Santiago de Compostela, Galicia-Spain}



\begin{abstract}
We perform a numerical analysis of the two particle azimuthal correlations at central rapidities generated in $p$A collisions within the  framework of the Color Glass Condensate. We extend the standard computations to include the subeikonal corrections which stem from considering the finite longitudinal width of the dense target.  For practical reasons, we only consider the next-to-next-to-eikonal corrections instead of using the all-order expressions for the inclusive two gluon production cross section. We show that the subeikonal terms that we account for in the two gluon yields contribute to both even and odd harmonics, the latter being absent in the standard Color Glass Condensate calculations performed at eikonal accuracy. Our analysis confirms the vanishing of the subeikonal effects with increasing collision energy and when going to forward rapidities, as expected. 
\end{abstract}

\end{frontmatter}
\section{Introduction}
\label{sec:introduction}


Experimental analyses of particle correlations performed in collision systems from large to small: heavy ions, proton-nucleus ($p$A) and proton-proton ($pp$), showed an asymmetry of the azimuthal distribution of the produced particles. Such asymmetry persists even for large rapidity difference of the particles, for which it has been named the ridge. This finding is among the key ones that demonstrate the existence of collective effects in small systems~\cite{Schlichting:2016sqo,Schenke:2017bog,Loizides:2016tew,Citron:2018lsq,Nagle:2018nvi}. These collective effects can be explained either by final state effects, where an anisotropic hydrodynamical flow is generated due to spatial asymmetries which result in momentum ones  (see e.g.~\cite{Jeon:2016uym,Romatschke:2017ejr}), or by initial state effects, where azimuthal correlations of the produced particles are seen as a result of the momentum correlations in the wave function of the colliding particles (see the review~\cite{Altinoluk:2020wpf}). Since the experiments involving azimuthal anisotropies probe the low-$x$ region of the nuclear target, the explanations based on the initial stage of the collision are usually performed within the Color Glass Condensate (CGC) effective theory~\cite{Gelis:2010nm,Kovchegov:2012mbw}.
	
	In the present manuscript we  focus on the azimuthal angle dependence of the two-gluon spectrum at central rapidities generated in a $p$A collision within the CGC. The two-particle azimuthal distribution of particles with transverse momenta ${\bf k}_1$ and ${\bf k}_2$ and rapidities $\eta_1$, $\eta_2$ is quantified through its Fourier harmonics that we define as
	\begin{align}\label{eq:azimuthal_harm}
		v_n^2(p_\perp, p_\perp^{\rm ref}) = \frac{\int\frac{d^2 N}{d\eta_1 d^2 {\bf k}_1 d\eta_2 d^2 {\bf k}_2} e^{i n (\phi_2-\phi_1)} d\phi_1 d\phi_2}{\int \frac{d^2 N}{d\eta_1 d^2 {\bf k}_1 d\eta_2d^2 {\bf k}_2} d\phi_1 d\phi_2} \Bigg|_{\substack{|{\bf k}_1|= p_\perp \\  |{\bf k}_2| = p_\perp^{\rm ref}}},
	\end{align}
	where $\phi_i$ is the azimuthal angle of the $i^{\rm th}$ particle, $d^2 {\bf k}_i=|{\bf k}_i| d|{\bf k}_i| d \phi_i$ and $\frac{d^2 N}{d\eta_1 d^2 {\bf k}_1 d\eta_2 d^2 {\bf k}_2}$ is the two-particle spectrum. Note that other definitions exist but the exact definition of the harmonics is not relevant for our purposes.
	
	In high energy collisions, where the small-$x$ tail of the projectile or target wave function is probed, the CGC offers a nonperturbative but weak coupling description of the system. In this regime where parton densities in the target are saturated, the projectile is assumed to be populated by almost collinear gluons that scatter eikonally on a dense and Lorentz contracted target. It is well known that in this scenario the scattering matrix for a gluon moving in the plus light-cone direction is described by a Wilson line in the adjoint representation,
	\begin{align}
	\mathcal{U}_{[x^+,y^+]}({\bf x}) = \mathcal{P}^+ \exp \left\{ -ig \int_{y^+}^{x^+} dz^+ A^-(z^+, {\bf x})\right\},
	\end{align}
	which represents the path ordered propagation of a gluon, from $y^+$ to $x^+$ at a fixed transverse position ${\bf x}$, in a classical field, $A^\mu(x) \equiv T^a A^\mu_a(x)$, dominated by its minus component.
	In standard CGC calculations, the reality of the adjoint representation implies that the leading order two-gluon spectrum has an accidental ${\bf k}_i \rightarrow - {\bf k}_i$ symmetry which prevents the generation of odd azimuthal harmonics (see e.g. the discussion in~\cite{Davy:2018hsl}). This absence of odd harmonics is, of course, not consistent with experimental results.
	
	Among several proposed solutions to this problem~\cite{Altinoluk:2020wpf}, it has been seen recently that relaxing the eikonal approximation by taking into account a finite target width, the scattering amplitude is no longer real and therefore the two-gluon azimuthal distribution entails odd harmonics~\cite{Agostini:2019avp,Agostini:2019hkj}. This property has been proven in the case of a dilute target, within the glasma graph approximation suitable for $pp$ collisions, where the scattering amplitude is modified by a non-eikonal phase $-i k^- L^+$ containing the target length along the plus light-cone direction $L^+$ and the minus light-cone momentum of the produced gluon $k^-$. 
		
	In this manuscript our aim is to generalize the numerical study of finite width target effects on two-particle azimuthal correlations performed in~\cite{Agostini:2019hkj} for $pp$ collisions to the case of $p$A collisions, where the target is modelled as a dense gluonic ensemble with a finite (light-cone) width $L^+$. The framework for including these corrections has been presented by us in a recent work~\cite{Agostini:2022ctk} where we have shown the equivalence to the standard approach for jet quenching (see the reviews~\cite{Casalderrey-Solana:2007knd,Mehtar-Tani:2013pia,Blaizot:2015lma}) where the final state gluon acquires its transverse momentum through a random walk inside the target medium. In this case, in contrast to the eikonal one, the scattering matrix is non-diagonal in transverse position and it is written in terms of the scalar gluon propagator,
	\begin{align}\label{eq6:scalar_propagator}
	\mathcal{G}_{k^+}(x^+,{\bf x} ; y^+,{\bf y}) = \int_{{\bf z}(y^+)={\bf y}}^{{\bf z}(x^+)={\bf x}} [\mathcal{D} {\bf z}(z^+)] 	
	\exp \left\{ \frac{i k^+}{2} \int_{y^+}^{x^+} dz^+ \dot{{\bf z}}^2 \right\} \mathcal{U}_{[x^+,y^+]}({\bf z}),
	\end{align}
	which represents the Brownian motion of a gluon with finite longitudinal momentum\footnote{Note that the eikonal approximation is recovered taking $k^+ \to \infty$.} $k^+$ inside a classical gauge field $A^-(z^+,{\bf z})$ from transverse position ${\bf y}$ at $y^+$ to ${\bf x}$ at $x^+$. The method for computing multi-gluon production, based on the McLerran-Venugopalan (MV) and Area Enhancement (AE) models, is given in~\cite{Agostini:2022ctk,Agostini:2021xca}. However, in this work we will restrict ourselves to the numerical analysis of two-gluon production. As usual in CGC calculations, we assume that correlations between the final particles reflect those of the produced gluons and are not washed out by final state effects or hadronisation. Our aim here is not to attempt any description of experimental data, as we assume that our model does not contain the full ingredients required for it, but to examine where the contribution from non-eikonal effects on odd harmonics is sizable and where the approximations that we employ break.
	
	This manuscript is structured as follows: In \cref{sec:2gluonNNE} we review very briefly the formalism developed in~\cite{Agostini:2022ctk}, with detailed formulae provided in~\ref{app:framework}. In \cref{sec:numerical_results} we describe the numerical results, while our conclusions and outlook are given in \cref{sec:conclu}.
\section{Two-gluon production beyond the shockwave approximation}
\label{sec:2gluonNNE}

	The building blocks of the multi-gluon spectrum in $p$A collisions within the eikonal approximation are colour neutral objects known as multipoles. In particular, in the case of two-gluon production the building blocks are the dipole function, $\tfrac{1}{N_c^2-1} \langle {\rm Tr}[\mathcal{U} \mathcal{U}^\dagger] \rangle$, the double dipole function, $\tfrac{1}{(N_c^2-1)^2} \langle {\rm Tr}[\mathcal{U} \mathcal{U}^\dagger]\,{\rm Tr}[\mathcal{U} \mathcal{U}^\dagger] \rangle$, and the quadrupole function, $\tfrac{1}{N_c^2-1} \langle {\rm Tr}[\mathcal{U} \mathcal{U}^\dagger \mathcal{U} \mathcal{U}^\dagger] \rangle$, where $N_c$ is the dimension of the colour gauge group and $\langle \cdots \rangle$ denotes the average over all possible configurations of colour charge density inside the target. Within the AE model~\cite{Agostini:2022ctk,Agostini:2021xca,Kovner:2017ssr,Kovner:2018vec,Altinoluk:2018ogz}, we can assume that the Wilson lines follow a Gaussian distribution and therefore, using the Wick theorem, the multipoles can be written in terms of dipole functions. 
	
	As discussed in~\cref{sec:introduction}, if we relax the eikonal approximation by including a finite width\footnote{Strictly speaking this is a relaxation of the shockwave approximation, $L^+ \to 0$, since a proper calculation beyond eikonality is more involved: it should also take into account the $x^-$ dependence~\cite{Altinoluk:2021lvu} and the perpendicular component~\cite{Altinoluk:2020oyd,Chirilli:2021lif} of the gluon field, classical quark fields in the target wave function, quantum corrections to the classical fields and non light-like components of the initial currents.} for the target, $L^+$, the scattering matrix -- or equivalently the scalar gluon propagator within a classical medium, $\mathcal{G}$ -- becomes a more involved object than a Wilson line. Thus, within the AE model, the non-eikonal multi-gluon spectrum will be described by three building blocks that generalize the dipole function for a finite width target. These objects are $\tfrac{1}{N_c^2-1}\langle {\rm Tr}[\mathcal{U} \mathcal{U}^\dagger] \rangle$, $\tfrac{1}{N_c^2-1}\langle {\rm Tr}[\mathcal{G} \mathcal{U}^\dagger] \rangle$ and $\tfrac{1}{N_c^2-1}\langle {\rm Tr}[\mathcal{G} \mathcal{G}^\dagger] \rangle$. As indicated in \cref{eq6:scalar_propagator}, the generalizations of the dipole function are given by path integrals. In~\cite{Agostini:2022ctk} we have computed these objects in the case of a small dipole, $|{\bf x}-{\bf y}| \ll \Lambda_{\rm QCD}^{-1}$, where the correlator of two Wilson lines can be written as a Gaussian function and the resulting harmonic oscillator-like path integrals can be solved analytically. In  \ref{app:framework} we summarize the results obtained within this approach.
	
	The non-eikonal two-gluon yield depends, apart from the transverse momenta ${\bf k}_i$ ($i=1,2$), on the longitudinal momenta $k_i^+$ of the produced particles, on the target longitudinal width $L^+$ (assuming that the gauge field has support within a region $z^+\in[0,L^+]$ of the longitudinal space) and on the saturation momentum $Q_s$, which stems from the application of the MV~\cite{McLerran:1993ni,McLerran:1993ka,McLerran:1994vd} model to the target ensemble.  Thus, it is natural to define the non-eikonal parameter\footnote{Note that the definition in this work is different than the one that we gave in~\cite{Agostini:2022ctk} where  the non-eikonal parameter read
	\begin{equation}\nonumber
		\epsilon_i = \sqrt{\frac{iQ_s^2 L^+}{2 k_i^+}}.
	\end{equation}
	Since it is real valued and does not involve a square root, the definition in \cref{eq:noneik_par} is more convenient for a numerical  or perturbative analysis.}
	\begin{equation}\label{eq:noneik_par}
	\epsilon_i = \frac{Q_s^2 L^+}{2 k_i^+}.
	\end{equation}
	When $k_i^+ \to \infty$ the non-eikonal parameter vanishes and we recover the eikonal result. Since the transverse momenta acquired by the produced gluons are of order $Q_s$, we can interpret $2 k_i^+/Q_s^2$ as their correlation length, $\lambda_i^+$. Thus we can write $\epsilon_i \sim L^+/\lambda_i^+$ so that the eikonal approximation is recovered when $\lambda_i^+ \gg L^+$ as we should expect. We also note that, as discussed in~\cite{Agostini:2022ctk}, corrections to the AE model start to be sizable when $\epsilon_i > 1$.
	
\begin{figure}[h!]
		\centering
		\includegraphics[width=\textwidth]{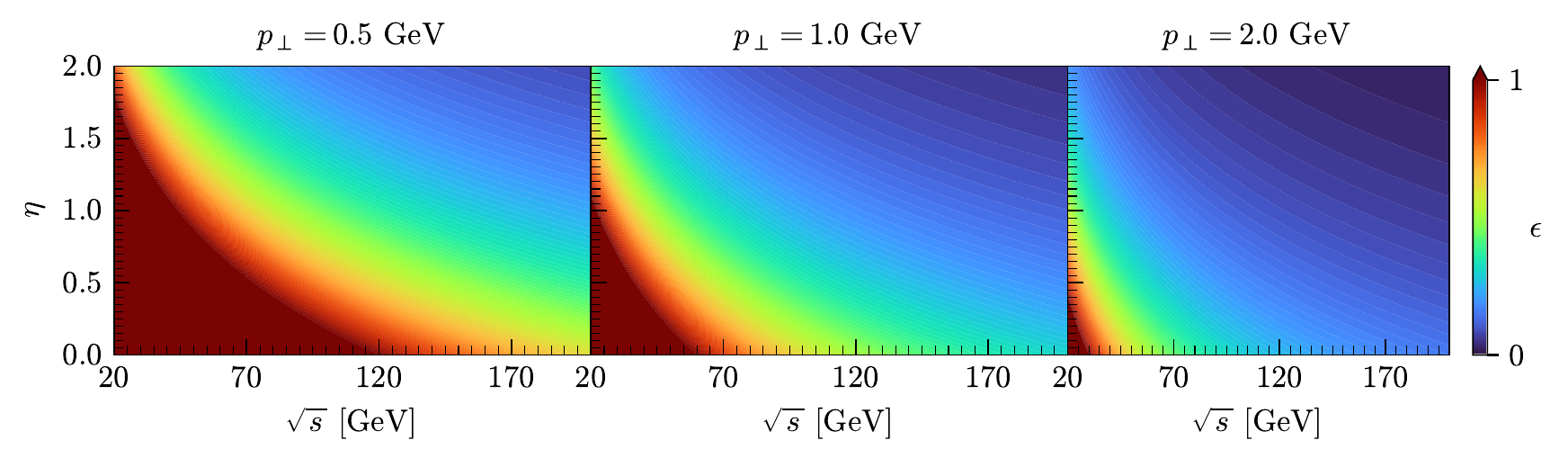}
		\caption{Dependence of the non-eikonal parameter, \cref{eq:noneik_par2}, on $p_\perp$, $\eta$ and $\sqrt{s}$ with a target defined by $Q_s= 1$ GeV and $A^{1/3}=6$. We see that the analysis performed in this work is valid, i.e., $\epsilon \lesssim 1$, when $p_\perp \gtrsim 1$ GeV and $\sqrt{s}\gtrsim 50$ GeV.}
		\label{fig:noneik_par}
	\end{figure}

	If we work in the center-of-mass (CoM) frame, the non-eikonal parameter can be written in terms of the collision CoM energy $\sqrt{s}$ and the pseudorapidities $\eta_i$ of the produced gluons. This representation in terms of measured quantities is more convenient for phenomenological analyses. In order to do so we note that, in the CoM frame, the Lorentz-contracted target width can be written as $L^+=2 r_A/(\gamma \sqrt{2})$, where $r_A \sim 5 A^{1/3} \ {\rm GeV}^{-1}$ is the nuclear radius, $A$ the mass number, $\gamma=\sqrt{s}/{(2 m_N)}$ the Lorentz factor and $m_N \sim 1$ GeV the nucleon mass. On the other hand, the longitudinal momentum of the on-shell produced gluons can be expressed in terms of their pseudorapidities as $k_i^+=|{\bf k}_i| e^{\eta_i}/\sqrt{2}$. Thus we can write
	\begin{equation}\label{eq:noneik_par2}
	\epsilon_i \sim 10 \frac{Q_s^2 A^{1/3}}{|{\bf k}_i| \sqrt{s}} e^{-\eta_i}.
	\end{equation}
	In \cref{fig:noneik_par} we plot this parameter as a function of $\sqrt{s}$ and $\eta$ for three values of the gluon transverse momentum $p_\perp \equiv |{\bf k}|$, where we have chosen a target with $Q_s=1$ GeV and $A^{1/3}=6$. We see that, for $p_\perp \gtrsim 1 $ GeV, $\epsilon \lesssim 1$ in most of the kinematic region probed at the Relativistic Heavy Ion Collider (RHIC), $\sqrt{s} \in [20, 200]$ GeV and $\eta \in [-4,4]$, and the Electron Ion Collider (EIC)\footnote{While the EIC will study electroproduction on protons and nuclei rather than $p$A collisions,  we expect that the general conclusions extracted here hold there, and an analysis similar to the one in this work can be done in DIS processes.}, $\sqrt{s} \in [30, 90]$ GeV and $\eta \in [-4,4]$. However, when $p_\perp \sim 0.5$ GeV we have that $\epsilon \sim 1$ in a large region of the kinematic space and the present analysis would not be justifiable. From \cref{fig:noneik_par} we  conclude that our analysis should be valid at momenta $p_\perp \gtrsim 1 $ GeV and CoM energies $\sqrt{s} \gtrsim 50$ GeV.
	
	The two-gluon spectrum given in \cref{eq:double_gluon_production} has a complicated non linear dependence on the parameters $\epsilon_1$ and $\epsilon_2$ which makes the evaluation of the azimuthal harmonics, \cref{eq:azimuthal_harm}, computationally expensive. In order to make the numerical analysis simpler we have decided to expand \cref{eq:double_gluon_production} perturbatively in terms of $\epsilon_i$. This expansion is justified as long as $\epsilon_i \ll 1$ which is the case in the kinematic region probed in this work.
	
	Thus we expand \cref{eq:double_gluon_production} at the first non trivial order\footnote{The first order term in the expansion in $\epsilon_i$ is known~\cite{Altinoluk:2014oxa,Altinoluk:2015gia} to give a vanishing result for the single inclusive spectra; the same happens for the double inclusive. This observation is only true for the non-eikonal contributions that stem from finite size corrections, which are the ones considered in this work.}, i.e., at second order in $\epsilon_i$. We refer to this approximation as the Next-to-Next-to-Eikonal (NNE) order and it reads
		\begin{align}\label{eq:NNE_expansion}
	N_2(\epsilon_1,\epsilon_2)\big|_{\rm NNE} = N_2(0,0) + \frac{\epsilon_1^2}{2} \frac{\partial^2 N_2(\epsilon_1,0)}{\partial^2 \epsilon_1} \Bigg|_{\epsilon_1=0}
	+ \frac{\epsilon_2^2}{2} \frac{\partial^2 N_2(0,\epsilon_2)}{\partial^2 \epsilon_2}\Bigg|_{\epsilon_2=0}
	+ \epsilon_1 \epsilon_2 \frac{\partial N_2(\epsilon_1,\epsilon_2)}{\partial \epsilon_1 \partial \epsilon_2}\Bigg|_{\epsilon_1=\epsilon_2=0},
	\end{align}
	where we have introduced the shorthand notation $N_2(\epsilon_1,\epsilon_2) \equiv \tfrac{d^2N}{dk_1^+ d^2{\bf k}_1 dk_2^+ d^2 {\bf k}_2}$. 
	
\begin{figure}[h!]
		\centering
		\includegraphics[width=\textwidth]{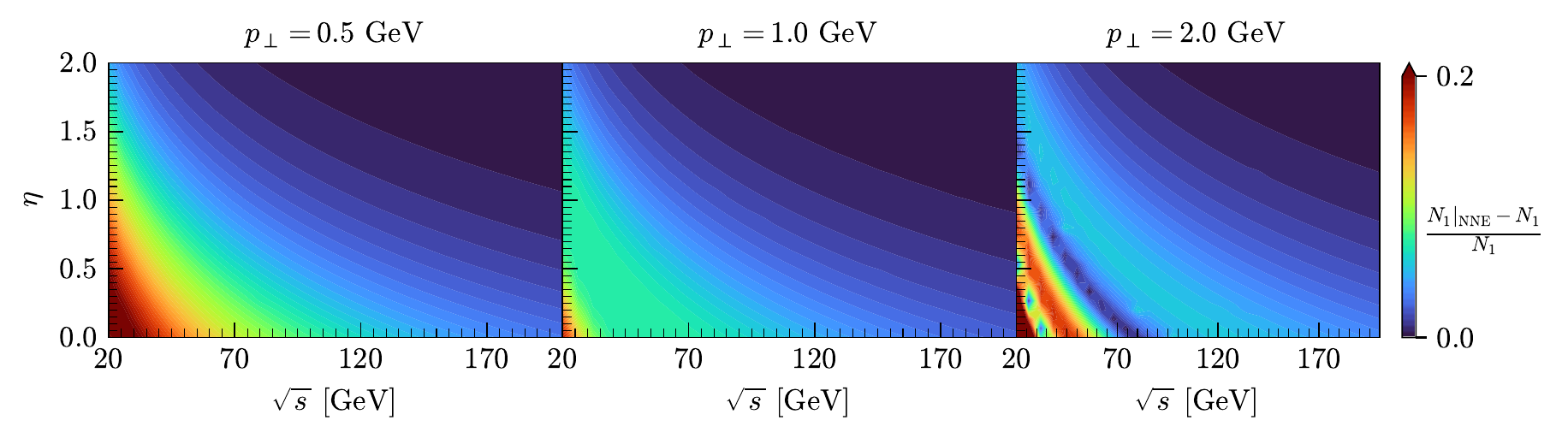}
		\caption{Relative difference between the NNE expansion of the single particle spectrum and its full non-eikonal result as a function of $p_\perp$, $\sqrt{s}$ and $\eta$, at a fixed value of $Q_s=1$ GeV and $A^{1/3}=6$. The oscillations in the difference in the plot on the right come from the oscillatory nature of the full result in some kinematic regions.}
		\label{fig:noneik_NNE}
	\end{figure}

	In order to double check the validity of the NNE expansion, in \cref{fig:noneik_NNE} we plot the relative difference of the single particle spectrum\footnote{Ideally we should study the relative difference of the NNE expansion of $N_2(\epsilon_1,\epsilon_2)$ with the full result. However, such analysis is more involved numerically. We expect a similar convergence for the case of two-gluon production since in \cref{eq:double_gluon_production} the dominant term is the single gluon spectrum squared while the other ones are suppressed by powers of $N_c^2-1$.} at NNE accuracy, $N_1(\epsilon)|_{\rm NNE}$, with the full non-eikonal result at different values of $p_\perp$, $\sqrt{s}$ and $\eta$ with a target defined by $Q_s=1$ GeV and $A^{1/3}=6$. We observe a better convergence than the one predicted by \cref{fig:noneik_par} since the relative difference is smaller than 20 \% in all the kinematic region studied in this work.
	
\section{Numerical results}
\label{sec:numerical_results}

	In this section we present the numerical results for the azimuthal harmonics, \cref{eq:azimuthal_harm}, extracted from the double gluon spectrum at NNE accuracy, \cref{eq:NNE_expansion}, for different values of $\epsilon_1$ and $\epsilon_2$. We fix $p_\perp^{\rm ref}=1$ GeV in the whole analysis.
	
	Before evaluating \cref{eq:NNE_expansion}, we have to deal with the infrared (IR) divergences that appear in the factors of \cref{eq:2point_target_aux}. In order to do so we introduce an IR regulator, $m_g=0.4$ GeV, in the denominators by making the change $1/{\bf q}_i^2 \to 1/({\bf q}_i^2+m_g^2)$ in \cref{eq:2point_target_aux}. We have checked that the result does not depend sizably under small variations of the parameter $m_g$. 
	
\begin{figure}[h!]
		\centering
		\includegraphics[width=\textwidth]{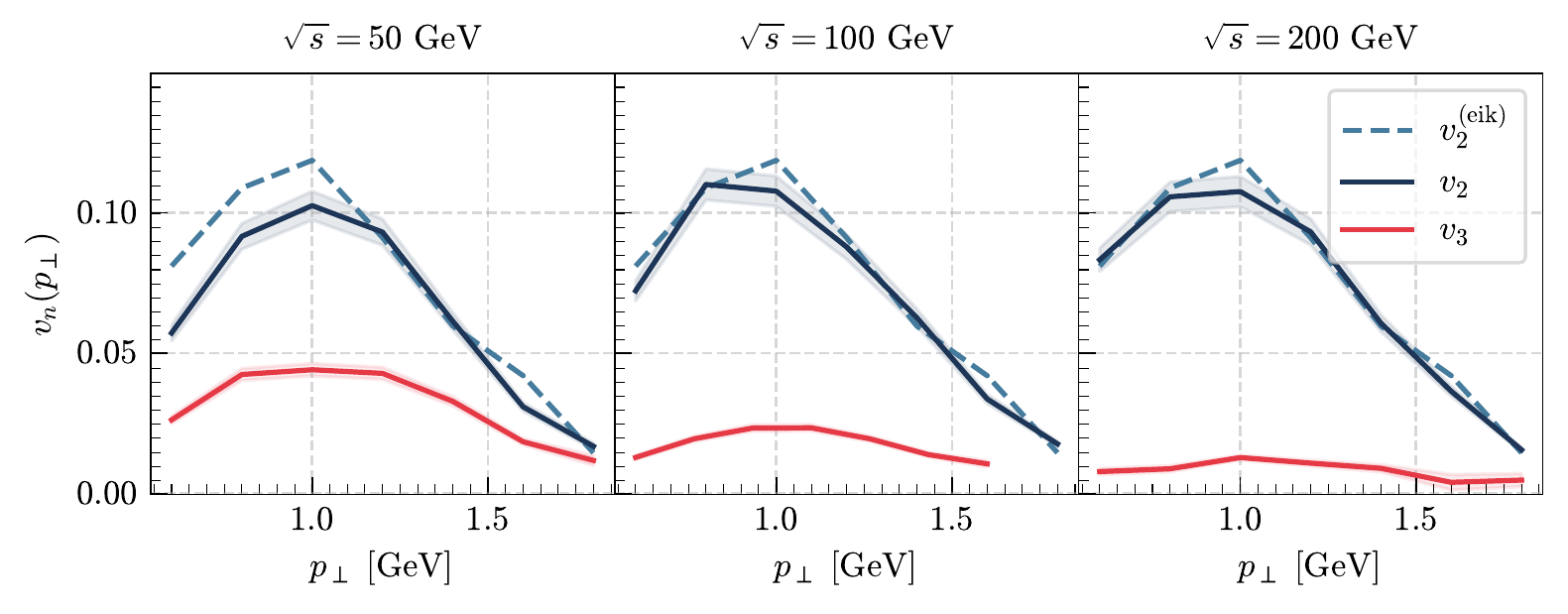}
		\caption{Dependence of the azimuthal harmonics, $v_2$ and $v_3$ on the transverse momentum at three values of the CoM energy. The bands indicate the estimated numerical error of the integration. For this plot we have fixed $\eta_1=0.1$, $\eta_2=0.5$, $Q_s=1$ GeV and $A^{1/3}=6$. We also compare the result with the one obtained in the eikonal approximation. In contrast to the eikonal case, we obtain a non vanishing $v_3$ that decreases as we increase the CoM energy of the collision. In the plot in the middle the point corresponding to $v_3$ at $p_\perp=1.8$ GeV is not shown as its numerical uncertainty is too large for it to be meaningful.}
		\label{fig:vn_pt}
	\end{figure}

	In \cref{fig:vn_pt} we plot the azimuthal harmonics $v_2$ and $v_3$ computed from \cref{eq:NNE_expansion} within a range $0.5 \ {\rm GeV}< p_\perp < 2 \ {\rm GeV}$ for three values of $\sqrt{s}$ and at fixed pseudorapidities $\eta_1=0.1$ and $\eta_2=0.5$. We have chosen a nucleus with $Q_s=1$ GeV and $A^{1/3}=6$. We also compare the result with the one that we get in the eikonal approximation ($\epsilon_1=\epsilon_2=0$) where, because of the accidental symmetry, $v_3=0$. We see that, in the non-eikonal case, we obtain a non vanishing odd azimuthal harmonic $v_3$ whose value becomes smaller at higher CoM energies as we should expect since we asymptotically approach the eikonal approximation. Besides, we should note that the result when $\sqrt{s}=50$ GeV has to be taken with care around the region with $p_\perp \lesssim 0.5$ GeV where the expansion parameter is not  small.
	
\begin{figure}[ht*]
		\centering
		\includegraphics[width=\textwidth]{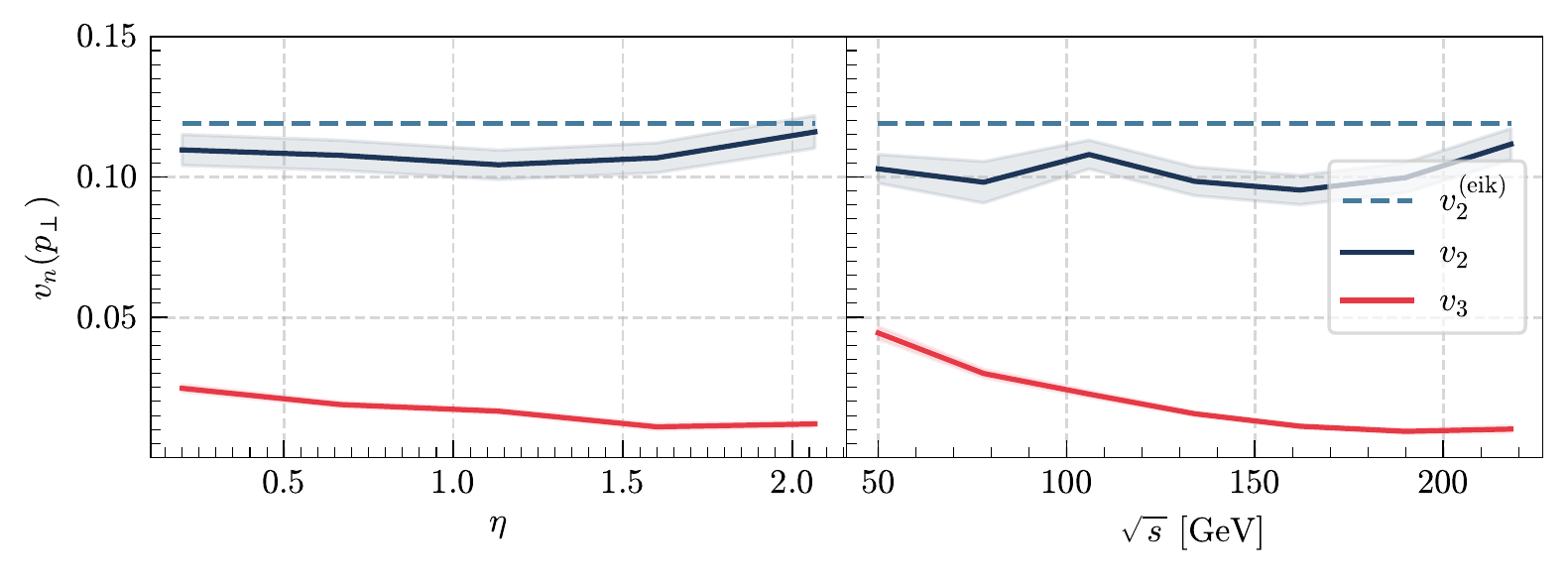}
		\caption{Left: dependence of the azimuthal harmonics on the pseudorapidity of one of the gluons $\eta \equiv \eta_1$ at a fixed value of $\sqrt{s}=70$ GeV and $\eta_2=1$. Right: dependence of the azimuthal harmonics on the CoM energy at a fixed value of $\eta_1=0.1$ and $\eta_2=0.5$. The bands indicate the estimated numerical error of the integration. In both plots we have fixed $p_\perp=1$ GeV, $Q_s=1$ GeV and $A^{1/3}=6$.}
		\label{fig:vn_eta}
	\end{figure}

	In \cref{fig:vn_eta} we plot the dependence of the azimuthal harmonics with the CoM energy at fixed $\eta_1=0.1$ and $\eta_2=0.5$, and the dependence with the pseudorapidity of one of the gluons at fixed $\sqrt{s}=70$ GeV  and $\eta_2=1$. Both plots show results computed with at a fixed value of $p_\perp = 1$ GeV and compared with the eikonal result. We see that as we increase the value of the pseudorapidity or the CoM energy, i.e., as we asymptotically approach the eikonal limit, the value of $v_3$ decreases and $v_2$ approaches $v_2^{\rm (eik)}$. We observe an unexpected behaviour of $v_2$, approaching the eikonal value, for small $\eta_1$ (left plot) or $\sqrt{s}$ (right plot) that we attribute to the increasing value of $\epsilon_i$ which deteriorates the validity of the NNE expansion in these kinematics.
	
	\section{Conclusions and outlook}
	\label{sec:conclu}
	
	In this work we have quantified the effect of the non-eikonal corrections that come from considering a target with a finite longitudinal size, on the azimuthal asymmetries of produced particles in $p$A collisions at central rapidities. We generalise the CGC expressions in order to include these corrections, as developed in~\cite{Agostini:2022ctk}, and check that in $p$A collisions odd azimuthal harmonics, absent in standard eikonal computations, are indeed present. For practical reasons we do not consider the full non-eikonal expressions but truncate them to next-to-next-to eikonal accuracy, which is the lowest order that provides non-trivial results. We check the validity of such truncation and provide the final results for the second and third order azimuthal harmonics. We observe the vanishing of the non-eikonal effects with increasing collision energy and when going to forward rapidities, as expected and observed in previous works for $pp$ collisions~\cite{Agostini:2019hkj}.
	
	We note that, as usual in CGC calculations, we assume that correlations between the final particles reflect those of the produced gluons and are not washed out by final state effects or hadronisation. We do not to attempt any description of experimental data, as we assume that our model does not contain the full ingredients required for it.
	
	The numerical analysis in this work is performed for $p$A collisions and therefore should be applicable to the low energy regime of the Relativistic Heavy Ion Collider. As an immediate outlook of this work, we plan to perform this analysis in $e$A collisions in order to check that we get similar results in Deep Inelastic Scattering at the energies of the Electron Ion Collider, even if only part of the non-eikonal corrections are included in the framework presented here.

		
Although conceptually more challenging, two other directions must be explored. On the one hand, so far we have only included the corrections due to a finite width of the target. The next step is to use the results of~\cite{Altinoluk:2021lvu,Altinoluk:2020oyd} and generalize this analysis in order to include  $A_\perp$ and the dependence of the gauge field on $x^-$. In order to do this we must modify the standard MV model to include averages $\langle A^\mu(x) A^\nu(y) \rangle$ for all components of the gauge field that depend on all four coordinates. On the other hand, we must study the limitations to the AE model that are already present at eikonal accuracy and are expected to be larger beyond the eikonal approximation. 	

\section*{Acknowledgements}
PA and NA have received financial support from Xunta de Galicia (Centro singular de investigaci\'on de Galicia accreditation 2019-2022), by European Union ERDF, by  the ``Mar\'{\i}a  de Maeztu" Units  of  Excellence program  MDM-2016-0692, and by the Spanish Research State Agency under project PID2020-119632GB-I00. TA is supported in part by the National Science Centre (Poland) under the research grant no. 2018/31/D/ST2/00666 (SONATA 14). PA is supported by grant ED481B-2022-050 by Xunta de Galicia.
This work has been performed in the framework of the European Research Council project ERC-2018-ADG-835105 YoctoLHC and the MSCA RISE 823947 ``Heavy ion collisions: collectivity and precision in saturation physics''  (HI\-EIC), and has received funding from the European Un\-ion's Horizon 2020 research and innovation programme under grant agreement No. 824093.

\appendix

\section{Double gluon production beyond the shockwave approximation}
\label{app:framework}

In this appendix we summarize the results of~\cite{Agostini:2022ctk} that are relevant for the present work. In~\cite{Agostini:2022ctk}, we have used the harmonic oscillator approximation in order to solve the generalized dipole functions within a finite width background field. The main results can be written\footnote{Note that there is a redefinition $\epsilon_i \to \sqrt{i \epsilon_i}$ with respect to~\cite{Agostini:2022ctk}.}
\begin{align}
	\label{eq:gbw_local}
	d^{(0)}(x^+,y^+|{\bf y},\bar{\bf y})
	\equiv \frac{1}{N_c^2-1} \Big\langle {\rm Tr}\Big[\mathcal{U}_{[x^+,y^+]}({\bf y}) \mathcal{U}^\dagger_{[x^+,y^+]}(\bar{\bf y})\Big] \Big\rangle 
	=\exp\Bigg\{- \frac{\Delta^+}{L^+}\frac{Q_s^2}{4} ({\bf y}-\bar{\bf y})^2\Bigg\},
\end{align}
\begin{align}
	\label{eq:gu_function}
	d^{(1)}(x^+,y^+|{\bf x},{\bf y},k_i^+;\bar{\bf y}) &\equiv
	\frac{1}{N_c^2-1} \Big\langle {\rm Tr} \Big[ \mathcal{G}_{k_i^+}(x^+,{\bf x} ; y^+,{\bf y}) \mathcal{U}^\dagger_{[x^+,y^+]}(\bar{\bf y}) \Big] \Big\rangle 
	\nonumber \\ & =
	\frac{- Q_s^2}{4 \pi \sqrt{i \epsilon_i} \sin\frac{\sqrt{i \epsilon_i} \Delta^+}{L^+}} \exp\Bigg\{\frac{Q_s^2}{4\sqrt{i \epsilon_i}} \left[ \frac{({\bf y}-{\bf \bar y})^2+({\bf x}-{\bf \bar y})^2}{\tan\frac{\sqrt{i \epsilon_i} \Delta^+}{L^+}} -2 \frac{ ({\bf y}-{\bf \bar y}) \cdot ({\bf x}-{\bf \bar y})}{\sin\frac{\sqrt{i \epsilon_i} \Delta^+}{L^+}} \right]\Bigg\}
\end{align}
and
\begin{align}
	\label{eq:gg_function}
	&d^{(2)}(x^+,y^+|{\bf x},{\bf y},k_1^+; \bar{\bf x},\bar{\bf y},k_2^+) \equiv \frac{1}{N_c^2-1}\Big\langle {\rm Tr} \Big[\mathcal{G}_{k_1^+}(x^+,{\bf x} ; y^+,{\bf y}) \mathcal{G}_{k_2^+}(x^+,\bar{{\bf x}} ; y^+,\bar{{\bf y}}) \Big] \Big\rangle
	\nonumber \\ & = \hskip0cm
	\frac{Q_s^4}{(4\pi)^2}
	\frac{\epsilon_{-} L^+ }{\Delta^+ \epsilon_1 \epsilon_2 \sin \frac{\Delta^+ \epsilon_{-}}{L^+}}
	\nonumber \\ & \hskip2cm \times
	{\rm exp} \Bigg\{\frac{Q_s^2}{4 \epsilon_{-}^2} \Bigg(
	\frac{\epsilon_{-} ({\bf r}_0^2+{\bf r}_N^2)}{\tan\frac{\Delta^+ \epsilon_{-}}{L^+}}
	-2\frac{\epsilon_{-} {\bf r}_0 \cdot {\bf r_N}}{\sin\frac{\Delta^+ \epsilon_{-}}{L^+}}
	+
	\frac{L^+}{\Delta^+ \epsilon_1 \epsilon_2 \epsilon_{+}^4} \left[
	2\epsilon_1 \epsilon_2 ({\bf r}_N-{\bf r}_0)
	+\epsilon_{+}^2 \epsilon_{-}^2
	({\bf b}_N-{\bf b}_0)
	\right]^2
	\Bigg)	
	\Bigg\}.
\end{align}
These equations were computed for $x^+>y^+$ and $\Delta^+ = [\min(x^+,L^+)-\max(y^+,0)] \Theta(x^+) \Theta(L^+-y^+)$ is the longitudinal width traversed by the emitted gluon inside the medium, $\epsilon_{\pm}^2=i(\epsilon_1\pm\epsilon_2)$, ${\bf r}_0={\bf y}-{\bf \bar{y}}$, ${\bf r}_N={\bf x}-{\bf \bar{x}}$, ${\bf b}_0=(k_1^+ {\bf y}+k_2^+ {\bf \bar{y}})/(k_1^++k_2^+)$ and ${\bf b}_N=(k_1^+ {\bf x}+k_2^+ {\bf \bar{x}})/(k_1^++k_2^+)$. 

The double inclusive gluon spectrum can be written in terms of the generalized dipole functions by using the Area Enhancement model which leads to the following expression:
\begin{align}
	\label{eq:double_gluon_production}
	\frac{d^2N}{d\eta_1 d^2{\bf k}_1 d\eta_3 d^2 {\bf k}_3} = \frac{g^4}{4(2\pi)^6} \int_{{\bf q}_1,{\bf q}_2,{\bf q}_3,{\bf q}_4}
	&\Bigg( \Big \langle \Omega_1 \Omega_2 \Big \rangle_{p} \Big \langle \Omega_3 \Omega_4 \Big \rangle_{p}
	+ \Big \langle \Omega_1 \Omega_3 \Big \rangle_{p} \Big \langle \Omega_2 \Omega_4 \Big \rangle_{p}
	+ \Big \langle \Omega_1 \Omega_4 \Big \rangle_{p} \Big \langle \Omega_2 \Omega_3 \Big \rangle_{p}   \Bigg)
	\nonumber \\ \times &
	\Bigg(
	\Big \langle \Lambda_1 \Lambda_2 \Big \rangle_{T} \Big \langle \Lambda_3 \Lambda_4 \Big \rangle_{T}
	+ \Big \langle \Lambda_1 \Lambda_3 \Big \rangle_{T} \Big \langle \Lambda_2 \Lambda_4 \Big \rangle_{T}
	+ \Big \langle \Lambda_1 \Lambda_4 \Big \rangle_{T} \Big \langle \Lambda_2 \Lambda_3 \Big \rangle_{T}
	\Bigg),
\end{align}
where
\begin{align}
	\label{eq:projectile_correlator}
	\Big \langle \Omega_i \Omega_j \Big \rangle_{p}=\frac{\delta^{b_i b_j}}{N_c^2-1} \mu^2\big[(-1)^{i+1}\textbf{q}_i,(-1)^{j+1}\textbf{q}_j\big], \quad \mu^2[{\bf k},{\bf q}]= \pi B_p e^{-\frac{({\bf k}+{\bf q})^2}{4}B_p}
\end{align}
is the projectile charge density correlator and $B_p$ is the gluonic transverse area of the projectile which we have fixed to $B_p = 4$ GeV$^{-2}$ in the whole manuscript. On the other hand,
\begin{align}
	\label{eq:2point_target_aux}
	&\Big \langle \Lambda_\alpha \Lambda_\beta \Big \rangle_T
	=\frac{\delta^{a_\alpha a_\beta} \delta^{b_\alpha b_\beta}}{N_c^2-1} 
	\int_{\textbf{y}_\alpha, \textbf{y}_\beta, \textbf{x}_\alpha, \textbf{x}_\beta }
	e^{ i (-1)^{\alpha+1} (\textbf{q}_\alpha \cdot \textbf{y}_\alpha-\textbf{k}_\alpha \cdot \textbf{x}_\alpha) +i (-1)^{\beta+1} (\textbf{q}_\beta \cdot \textbf{y}_\beta-\textbf{k}_\beta \cdot \textbf{x}_\beta)}
	\nonumber \\ & \hskip 1cm \times
	\Bigg \{ 
	2\frac{{\bf k}_\alpha^{\lambda_\alpha}{\bf k}_\beta^{\lambda_\beta}}{{\bf k}_\alpha^{2}{\bf k}_\beta^{2}}
	\delta^{(2)}({\bf x}_\alpha-{\bf y}_\alpha) \delta^{(2)}({\bf x}_\beta-{\bf y}_\beta)
	d^{(0)}(L^+,0|{\bf y}_\alpha,{\bf y}_\beta)
	\nonumber \\ & \hskip 1cm
	-
	4\frac{{\bf k}_\alpha^{\lambda_\alpha}{\bf q}_\beta^{\lambda_\beta}}{{\bf k}_\alpha^{2}{\bf q}_\beta^{2}}
	\delta^{(2)}({\bf x}_\alpha-{\bf y}_\alpha)
	d^{(1)}(L^+,0|{\bf y}_\alpha;{\bf x}_\beta,{\bf y}_\beta,(-1)^{\beta+1}k_\beta^+)
	\nonumber \\ & \hskip 1cm
	+
	2\frac{{\bf q}_\alpha^{\lambda_\alpha}{\bf q}_\beta^{\lambda_\beta}}{{\bf q}_\alpha^{2}{\bf q}_\beta^{2}}
	d^{(2)}(L^+,0|{\bf x}_\alpha,{\bf y}_\alpha,(-1)^{\alpha+1}k_\alpha^+;{\bf x}_\beta,{\bf y}_\beta,(-1)^\beta k_\beta^+)
	\nonumber \\ & \hskip 1cm
	+
	2\frac{L^+}{k_\beta^+} \frac{{\bf k}_\alpha^{\lambda_\alpha}}{{\bf k}_\alpha^{2}}
	\delta^{(2)}({\bf x}_\alpha-{\bf y}_\alpha)
	\int_0^1 df_\beta 
	\Big[ \partial_{{\bf y}_\alpha^{\lambda_\alpha}} d^{(1)}(L^+, f_\beta L^+ | {\bf y}_\alpha;{\bf x}_\beta,{\bf y}_\beta,(-1)^{\beta+1}k_\beta^+) \Big]
	d^{(0)}(f_\beta L^+,0|{\bf y}_\alpha,{\bf y}_\beta)
	\nonumber \\ & \hskip 0cm
	-
	2\frac{L^+}{k_\beta^+} \frac{{\bf q}_\alpha^{\lambda_\alpha}}{{\bf q}_\alpha^{2}}
	\int_0^1 df_\beta  \int_{{\bf u}}
	\Big[ \partial_{{\bf y}_\beta^{\lambda_\beta}} d^{(2)}(L^+,f_\beta L^+|{\bf x}_\alpha,{\bf u},(-1)^{\alpha+1}k_\alpha^+;{\bf x}_\beta,{\bf y}_\beta,(-1)^\beta k_\beta^+) \Big]
	d^{(1)}(f_\beta L^+, 0 | {\bf y}_\beta;{\bf u},{\bf y}_\alpha,(-1)^{\alpha+1}k_\alpha^+)
	\nonumber \\ & \hskip 0.5cm
	+
	\frac{L^+}{k_\alpha^+} \frac{L^+}{k_\beta^+} 
	\int_0^1 df_\beta \int_0^{f_\beta} df_\alpha  \int_{{\bf u}}
	\Big[ \partial_{{\bf y}_\beta^{\lambda_\beta}} d^{(2)}(L^+,f_\beta L^+|{\bf x}_\alpha,{\bf u},(-1)^{\alpha+1} k_\alpha^+;{\bf x}_\beta,{\bf y}_\beta,(-1)^\beta k_\beta^+) \Big]
	\nonumber \\ & \hskip 3cm
	\times
	\Big[ \partial_{{\bf y}_\alpha^{\lambda_\alpha}} d^{(1)}(f_\beta L^+, f_\alpha L^+ | {\bf y}_\beta;{\bf u},{\bf y}_\alpha,(-1)^{\alpha+1}k_\alpha^+) \Big]
	d^{(0)}(f_\alpha L^+,0|{\bf y}_\alpha,{\bf y}_\beta)
	+(\alpha \leftrightarrow \beta)
	\Bigg\}
\end{align}
is the correlation function of the reduced matrix amplitude~\cite{Agostini:2022ctk} which contains the information about the target.

\vskip 1cm
\vskip 1cm
\bibliographystyle{elsarticle-num} 

\bibliography{mybib}

\end{document}